\documentclass[twocolumn,nolinenumbers]{aastex631}
\usepackage{physics}
\usepackage{hyperref}
\usepackage[varg]{txfonts}

\begin{document}
\title[]{Waltzing binaries: Probing line-of-sight acceleration of merging compact objects with gravitational waves}

\author{Aditya Vijaykumar \href{https://orcid.org/0000-0002-4103-0666}{\includegraphics[scale=0.03]{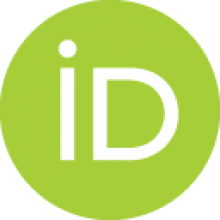}},$^{1,2}$
Avinash Tiwari \href{https://orcid.org/0000-0001-7197-8899}{\includegraphics[scale=0.03]{orcid-ID.png}},$^{3}$
Shasvath J. Kapadia  \href{https://orcid.org/0000-0001-5318-1253}{\includegraphics[scale=0.03]{orcid-ID.png}},$^{1, 3}$  K. G. Arun \href{https://orcid.org/0000-0002-6960-8538}{\includegraphics[scale=0.03]{orcid-ID.png}},$^{4}$ and Parameswaran Ajith \href{https://orcid.org/0000-0001-7519-2439}{\includegraphics[scale=0.03]{orcid-ID.png}}, $^{1,5}$}

\address{$^{1}$ International Centre for Theoretical Sciences, Tata Institute of Fundamental Research, Bangalore 560089, India\\
$^{2}$ Department of Physics, The University of Chicago, 5640 South Ellis Avenue, Chicago, Illinois 60637, USA\\
$^{3}$ The Inter-University Centre for Astronomy and Astrophysics, Post Bag 4, Ganeshkhind, Pune 411007, India \\
$^{4}$ Chennai Mathematical Institute, Siruseri, 603103 Tamilnadu, India \\
$^{5}$ Canadian Institute for Advanced Research, CIFAR Azrieli Global Scholar, MaRS Centre, West Tower, 661 University Ave, Toronto, ON M5G 1M1, Canada
}

\begin{abstract}
  Line-of-sight acceleration of a compact binary coalescence (CBC) event would modulate the shape of the gravitational waves (GWs) it produces with respect to the corresponding non-accelerated CBC. Such modulations could be indicative of its astrophysical environment.
  We investigate the prospects of detecting this acceleration in future observing runs of the LIGO-Virgo-KAGRA network, as well as in next-generation (XG) detectors and the proposed DECIGO. We place the first observational constraints on this acceleration, for putative binary neutron star mergers GW170817 and GW190425. We find no evidence of line-of-sight acceleration in these events at $90\%$ confidence.  Prospective constraints for the fifth observing run of the LIGO at A+ sensitivity, 
  suggest that accelerations for typical BNSs could be constrained with a precision of $a/c \sim 10^{-7}~[\mathrm{s}^{-1}]$, assuming a signal-to-noise ratio of $10$. These improve to  $a/c \sim 10^{-9}~[\mathrm{s}^{-1}]$ in XG detectors, and $a/c \sim 10^{-16}~[\mathrm{s}^{-1}]$ in DECIGO. We also interpret these constraints in the context of mergers around supermassive black holes.
\end{abstract}

\section{Introduction}\label{sec:introduction}
The LIGO-Virgo \citep{LIGOScientific:2014pky, VIRGO:2014yos} network of gravitational-wave (GW) detectors has observed $\sim 90$ GW events across three observing runs, all of which were produced by compact binary coalescences (CBCs) \citep{LIGOScientific:2021djp}. The next observing run promises to observe even more GW events, likely more than tripling the existing list of confirmed detections \citep{KAGRA:2013rdx}.

The vast majority of detected CBC events are merging binary black holes (BBHs) \citep{LIGOScientific:2021djp}. This offers the exciting prospect of constraining their population properties and exploring their formation channels \citep{LIGOScientific:2021psn}.  It has been suggested, however, that a single formation channel cannot explain all the detected BBH events and their source properties \citep{Zevin:2020gbd}.  Furthermore, it is often difficult to conclusively identify a given event's formation channel, although statistical arguments can sometimes be made to quantify if one formation channel is preferred over another. Nevertheless, even if such arguments can point to one channel being favored over another, they cannot always conclusively rule out all formation channels at the exception of the statistically preferred one. Binary neutron star (BNS)~\citep{GW170817-DETECTION,GW190425-DETECTION} and neutron star black hole binaries (NSBH)~\citep{NSBH-DETECTION} have also been observed with GWs. However, given that their number is barely a handful \citep{LIGOScientific:2021djp}, it is difficult to even make statistical arguments about their provenance, although some preliminary constraints on their source properties have been placed. 

It is therefore of considerable interest to ask if there exist any potential signatures or ``smoking guns'', in the GW waveform itself, that could help identify the astrophysical environment or the formation channel of the CBC that produced it.  In this work, we discuss one such generic feature, the signature of accelerated motion of the binary in the gravitational waveform.

Non-rectilinear motion of the compact binary's center-of-mass is expected when the merger happens in a gravitational potential. Though there may be many scenarios where such mergers occur, the resulting motion will depend on the distance of the binary from the center of the potential. Here we focus on CBCs in the vicinity of supermassive BHs, and discuss the detectability of the resulting accelerated motion with the current and future generation GW detectors as a function of the mass of the SMBH. 
CBCs in active galactic nuclei (AGN)~\citep{AGNformation93,AGNformation2000,AGNformation2014} would be an example of this binary population, though our method would be sensitive to any mechanism which provides the binary a detectable line-of-sight acceleration.

Accelerated motion of the centre-of-mass of the CBC, with a non-zero time-varying velocity component along the line-of-sight, would produce a time-varying Doppler shift. This in turn would modulate the inspiral waveform with respect to its standard shape\footnote{It is intuitively straightforward to see that these modulations would be identical to those produced by a time-varying gravitational constant $G$ \citep{Yunes:2009bv}. However, this is not expected to occur on timescales comparable to the duration of the CBC in-band, if at all \citep[see, e.g][]{Vijaykumar:2020nzc}.}.   Previous work (\cite{Yunes:2010sm}, \cite{Bonvin:2016qxr}) showed that a constant line-of-sight acceleration would introduce a term in the post-Newtonian expansion of the phase at the $-4$PN order.  {``Waltzing CBCs''\footnote{The motion of the centre of mass of inspiralling binaries orbiting SMBHs, is evocative of a ``Waltz'' where dance partners orbit each other while their mutual centre of mass also traces out a larger orbit.}, i.e, CBCs orbiting SMBHs} will also introduce terms at lower post-Newtonian orders pertaining to higher time derivatives of the velocity (jerk, snap, etc). However, given the finite duration of CBC inspirals in-band, such effects are more difficult to measure, especially for ground-based detectors whose noise power-spectral densities \citep{LIGOScientific:2019hgc} rise sharply at frequencies below $\sim 10$ Hz due to seismic activity. 

In this work, we place the very first observational constraints on the line-of-sight acceleration of putative binary neutron star mergers GW170817 \citep{GW170817-DETECTION} and GW190425 \citep{GW190425-DETECTION}. We find no evidence of a line-of-sight acceleration. {This is consistent with current expectations that most BNSs evolve and merge in isolated environments, and are therefore expected to have accelerations much smaller than can be constrained with O2 and O3 data.}

We also study the prospects of constraining this acceleration for a range of component masses of the CBCs, in the fourth observing run of the LIGO-Virgo-KAGRA network \citep{KAGRA:2013rdx}, the A+ configuration of the LIGO detectors \citep{KAGRA:2013rdx}, next-generation (XG) detector configurations Cosmic Explorer and Einstein Telescope \citep{CE, ET}, and the space-based detector DECIGO \citep{Sato:2017dkf}. We also interpret these constraints in the context of mergers around SMBHs.

Throughout the paper, we will consider accelerations in units of the speed of light converted to SI units. ie. $ a/c\  [\mathrm{s}^{-1}]$. Unless otherwise specified, $ a $ is the acceleration of the source along 
the line-of-sight.

\section{Motivation and Methods}\label{sec:method}
A binary of total mass $ M_\mathrm{src} $ in the source frame will, in general, appear to have a total mass 
\begin{equation}\label{key}
 	M = M_\mathrm{src}(1 + z_\mathrm{cos}) (1 + z_\mathrm{dop}) 
\end{equation}
in the detector frame. Here, $ z_\mathrm{cos} $ is the cosmological redshift of the source, and $ z_\mathrm{dop} \approx v / c $ is the Doppler shift induced due to a (constant) line-of-sight velocity $ v $ of the source. In addition, if the binary also has a line-of-sight acceleration $ a/c $, the apparent detector frame mass would be
\begin{equation}\label{key}
    M_\mathrm{det} = M (1 + a/c \times t) \qq{.}
\end{equation}
The equation above assumes that $|z_\mathrm{dop}| \ll 1$ and acceleration is low $(|a/c| \times t \ll 1)$. As is evident, an accelerating source produces a time-varying detector-frame mass which leaves an imprint on the gravitational waveform. 
If $ \Psi_0(f) $ is the full GW phase without acceleration, and  $ \Psi(f) = \Psi_0(f) + \Delta \Psi (f) $ is the phase including the acceleration, \cite{Bonvin:2016qxr} showed that $ \Delta \Psi(f) $ is given by:
\begin{equation}\label{key}
 	\Delta \Psi (f) = \frac{25}{65536 \,\eta^2} ~ {\qty(\frac{G M}{c^3})} ~ \qty(\dfrac{a}{c}) ~ {v_f}^{-13} \qq{,}
\end{equation}
where $ v_f = (\pi G M f / c^3)^{1/3} $. {For a given value of the acceleration, the accumulated $\Delta \Psi$ across the bandwidth of a GW detector would be more for less massive systems. This means that, for audio-band {($\sim 10-1000$ Hz) GW} detectors, $ a /c $ would be best measured with BNSs or light BBHs.} \cite{Tamanini:2019usx} further derived corrections to the leading order term upto $ 1.5 $ PN order, and also forecasted constraints on $ a/c $ from stellar mass binaries with LISA.

We extend the calculation of $ \Delta \Psi(f) $ to include $ 3.5 $ PN corrections to the leading order and obtain
\begin{widetext}
\begin{multline}
	\label{eq:phase-perturbation}
    \Delta \Psi(f) = \frac{25}{{65536}\,\eta^2} ~ {\qty(\frac{G M}{c^3})}
  ~ \qty(\dfrac{a}{c}) ~ {v_f}^{-13}
 \Biggl[ 1 + \left(\frac{743}{126} + \frac{22}{3}\eta  \right) v_f^2  - \frac{64 \pi}{5} v_f^3 + \left( \frac{1755623}{84672} + \frac{32633}{756}\eta + \frac{367}{12}\eta^2 \right) v_f^4 \\ - \left( \frac{20807}{210} + \frac{574}{15}\eta \right) \pi v_f^5 +  \Biggl\{ - \frac{28907482848623}{35206617600} + \frac{9472}{75}\pi^2 + \frac{13696}{105}\gamma + \frac{13696}{105}\ln (4v_f) + \Biggl( \frac{3311653861}{1524096} \\ - \frac{451}{6}\pi^2 \Biggr)\eta  + \frac{2030687}{18144}\eta^2 + \frac{66287}{648}\eta^3 \Biggr\}v_f^6 - \Biggl( \frac{158992529}{317520} + \frac{1015907}{1890}\eta - \frac{419}{945}\eta^2 \Biggr)\pi v_f^7 \Biggr]\,,
\end{multline}
\end{widetext}
{where $\eta := m_1 m_2/M^2$ is the symmetric mass ratio of the binary.}
The full derivation of the Eq. \eqref{eq:phase-perturbation} is described in Appendix~\ref{sec:appendix}. 
{As we will show in Sec.~\ref{sec:forecasts-ground-based}, extending the calculation to higher PN orders is important to avoid systematics while inferring the acceleration.}

Astrophysically, a binary can have a non-zero acceleration due to multiple reasons. For example, such acceleration could arise due to a binary's circular orbit around an SMBH \citep{Inayoshi:2017hgw}. In general, for a spherically symmetric potential $\Phi(r)$, the acceleration $\mathbf{a}(r)$ is given by 
\citep{BinneyTremaine, BovyBook}
\begin{equation}
    \|{\mathbf{a}(r)\| = {\dv{\Phi(r)}{r}}}. 
\end{equation}
Specifically, for motion around an  SMBH, one can express the line-of-sight acceleration as
\begin{equation}\label{eq:physical_units} 
	a/c = \frac{\mathbf{a}(r) \cdot \hat{\mathbf{n}}}{c}  = 4.65 \times 10^{-12} \qty(\dfrac{M_\mathrm{BH}}{10^{10} M_\odot}) \qty(\dfrac{r}{1 \mathrm{pc}})^{-2} \cos \theta   \ \mathrm{s}^{-1}
\end{equation}
where $r$ is the distance from the centre of the potential, $M_\mathrm{BH}$ is the mass of the SMBH, and  $\theta $ is the angle that the acceleration vector makes with the line-of-sight vector $\hat{\mathbf{n}}$.
In our convention,
$ \cos \theta = 1 $ (ie. $ \theta = 0 $) means that the acceleration vector is pointed away from the observer.  Since the GW phase only allows for a measurement of $ a/c $, one can only constrain the quantity $ M_\mathrm{BH} \cos \theta / r^2 $ with GW observations.

\section{Results}\label{sec:results}
\begin{figure}[ht!]
	\centering 
	\includegraphics[width=0.85\columnwidth]{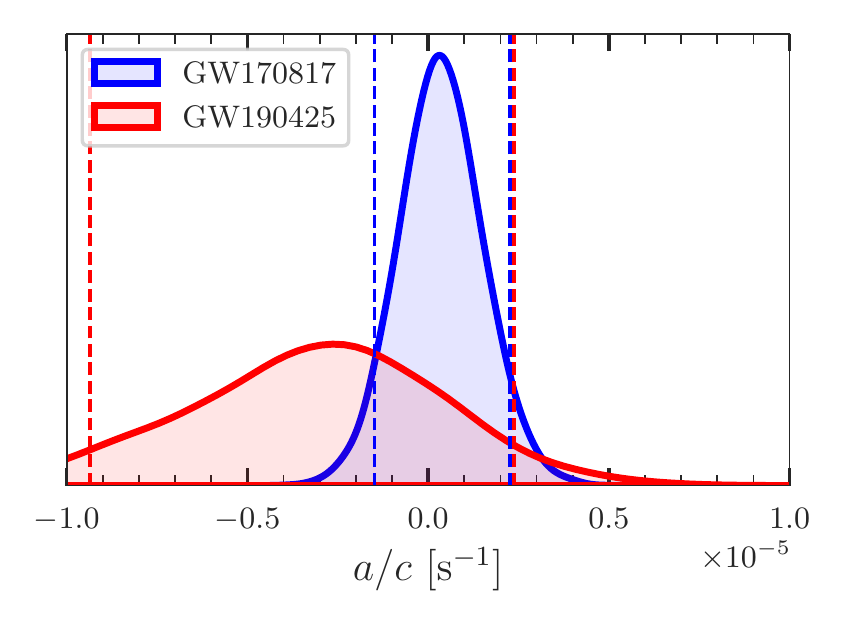}
	\caption{Measurement of the line-of-sight acceleration from GW170817 and GW190425. The measurements are expressed as a ratio of the acceleration to the speed of light ($a/c$) in units of $\mathrm{s}^{-1}$. The solid lines indicate the inferred posterior distribution on $a/c$, while the vertical dashed lines indicate the edges of the $90 \%$ CI.  {GW170817 yields a stronger constraint ($-1.5 \times 10^{-6}$---$2.2 \times 10^{-6} \ \mathrm{s}^{-1}$, $90\%$ CI) as compared to GW190425 ($-9.4 \times 10^{-6}$---$2.4 \times 10^{-6}\ \mathrm{s}^{-1}$, $90\%$ CI) due to its low detector frame chirp mass and also high SNR}. Both measurements are consistent with zero acceleration.}
	\label{fig:bns-constraints}
\end{figure}
\begin{figure*}[ht!]
	\centering 
	\includegraphics[width=0.85\linewidth]{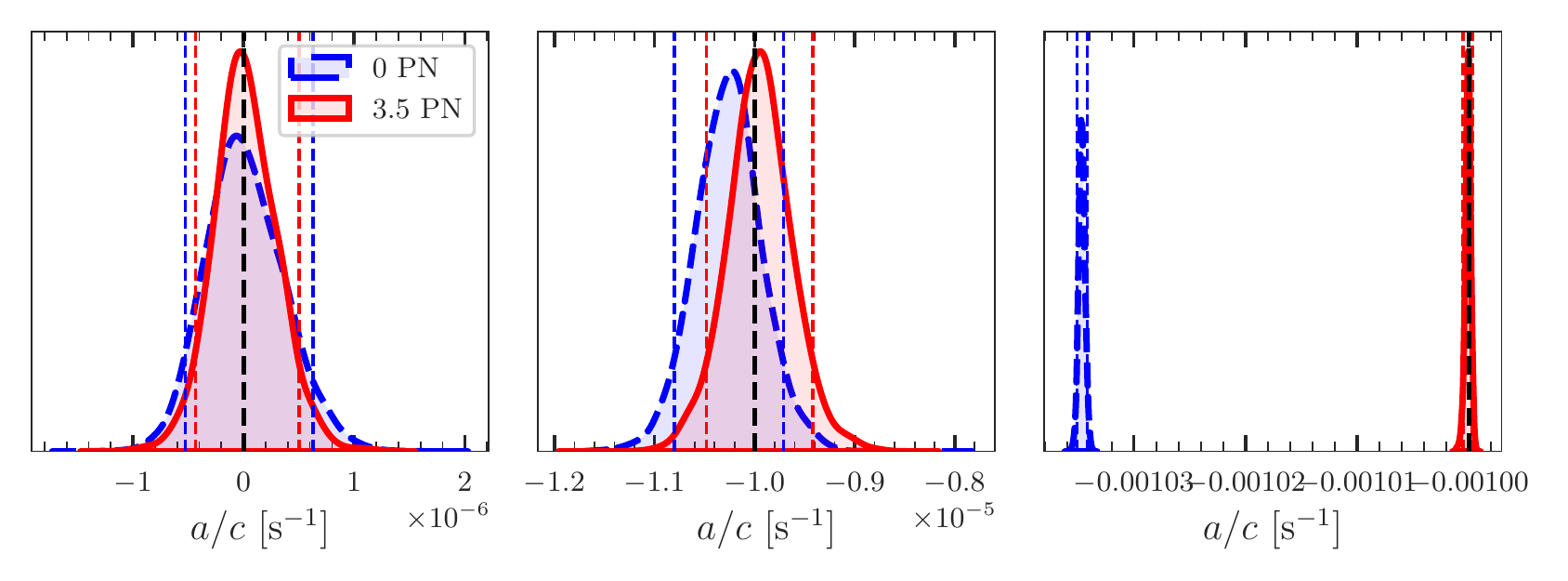}
	\includegraphics[width=0.85\linewidth]{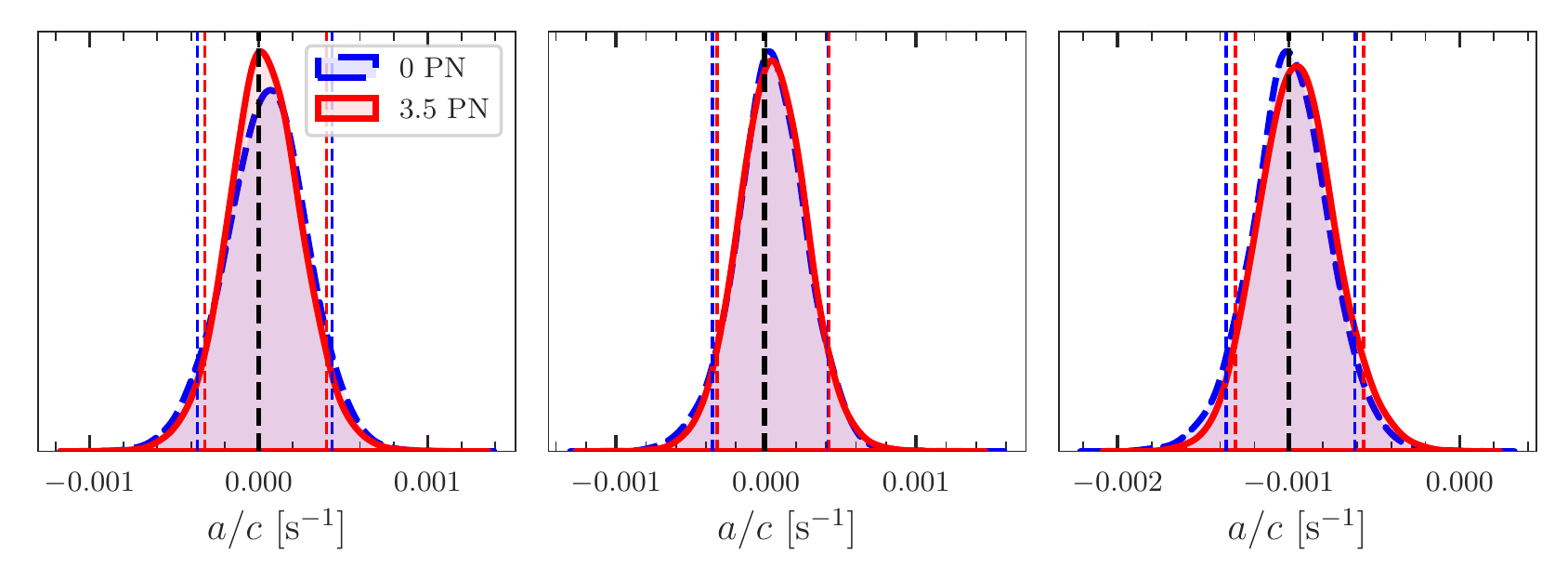}
	\caption{Recovered posteriors on $ a/c $ from GW170817-like (top) and GW170608-like (bottom)  injections in an O4 network. {The injected values (from left to right in each row) are $ 0 $, $ -10^{-5} $, and $- 10^{-3} $ s$^{-1}$.} These are plotted as vertical black dotted lines in each panel. The injected template contains the full phase of  Eq. \eqref{eq:phase-perturbation}. The recoveries are done with the full phase (red solid curve) as well as the leading-order phase to check (blue dashed curve) for systematics. The precision in recovery of $ a/c $ is $\sim10^{-7} \ \mathrm{s}^{-1}$ for GW170817, and $\sim 10^{-4}\ \mathrm{s}^{-1} $ for GW170608-like injections.  Using only the leading-order phase for recovery does cause  systematics in the recovery of the GW170817-like injections, with the bias being worse for larger injected values of $ a/c $. However, recoveries of the GW170608-like injections do not show any significant bias. 
    regardless of the injected $ a/c $.}
	\label{fig:O4-constraints}
\end{figure*}

\subsection{Constraints from GW170817 and GW190425}

We first measure the line-of-sight 
acceleration from binary neutron star candidates GW170817 \citep{GW170817-DETECTION} and GW190425 \citep{GW190425-DETECTION}. These events are chosen due to their low detector-frame chirp mass, which is ideally suited for a precise measurement of the acceleration. 
Template waveforms
for our analyses are constructed by adding phase corrections due to line-of-sight
acceleration (Eq.~\eqref{eq:phase-perturbation}) onto the {IMRPhenomPv2\_NRTidal} \citep{Dietrich:2018uni} approximant implemented within \texttt{lalsuite} \citep{lalsuite}. 
We perform Bayesian inference on these signals, using low-spin priors\footnote{We have verified that using high-spin priors doesn't qualitatively change our results, and only changes the width of the 90\% CI on $a/c$ by $\sim 10 \%.$} of \cite{LIGOScientific:2018hze} and \cite{GW190425-DETECTION}, sampling over all relevant intrinsic and extrinsic parameters including $a/c$ using the dynamic nested sampler \texttt{dynesty} \citep{Speagle:2019ivv}. 
The prior on $a/c$ is assumed to be flat between $-10^{-2} \ \mathrm{s}^{-1}$ and $10^{-2} \ \mathrm{s}^{-1}$.
We use the parameter estimation packages \texttt{bilby} \citep{Ashton:2018jfp} and \texttt{bilby\_pipe} \citep{Romero-Shaw:2020owr} for streamlining our analyses, while also using the relative binning/heterodyning scheme \citep{Cornish:2010kf, Zackay:2018qdy,2021PhRvD.104j4054C} to speed up our likelihood calculations \citep{Krishna:2022}. The likelihood is calculated in the range 20 Hz to 2048 Hz\footnote{Since the corrections due to acceleration are calculated using PN expressions, we should have ideally cut-off our analysis at a frequency beyond which these expressions aren't valid. However, we do not expect this choice to impact our results; the SNR is negligible at high frequencies, and the measurement of $a/c$ is driven by the low-frequency part of the signal.} assuming a sampling rate of 4096 Hz using the  publicly-released noise power spectral densities~\citep{LIGOScientific:2019lzm}, without marginalizing over calibration uncertainties. The inferred posterior on $ a/c $ for the two events is shown in Figure \ref{fig:bns-constraints}. {We find that both events yield a measurement consistent with zero line-of-sight 
acceleration, with the 90\% CI being $-1.5 \times 10^{-6}$---$2.2 \times 10^{-6} \ \mathrm{s}^{-1}$ for GW170817 and $-9.4 \times 10^{-6}$---$2.4 \times 10^{-6} \ \mathrm{s}^{-1}$ for GW190425. }

Assuming these binaries were orbiting a SMBH we now interpret these measurements as limits on the location of the binary around the SMBH. For this purpose, we directly use Eq.~\eqref{eq:physical_units}, and assume that the SMBH mass $M_\mathrm{BH}$ makes up most of the mass that is enclosed by the binary's orbit a distance $r$ away from the SMBH.
{In what follows, we quote constraints on $r$ marginalizing over $\theta$ assuming a uniform prior on $\cos \theta$. Since GW170817 also had an electromagnetic counterpart, the host galaxy of the merger was  confidently identified as NGC 4993 \citep{DES:2017kbs}, a galaxy that hosts a supermassive black hole of mass $ M_{BH} \approx 0.7 \times 10^8 M_\odot $ \citep{Levan:2017ubn}\footnote{This mass is calculated using the $M-\sigma$ relation of \cite{Gultekin:2009qn} along with NGC 4993's velocity dispersion estimate of $ \sigma \approx 170 \ \mathrm{km}  \ \mathrm{s}^{-1} $ \citep{Levan:2017ubn}.}.
For this event, we obtain a constraint $r > 12.1 \ \mathrm{AU}$ at the $90 \%$ credible level (CL). 
However, from the electromagnetic counterpart, we also know that GW170817 was $1.96  \ \mathrm{kpc}$ ($\approx 4 \times 10^8 \ \mathrm{AU}$) from the central black hole~\citep{Levan:2017ubn}. Hence, the constraints that we obtain for GW170817 are weaker by several orders of magnitude. Since the host galaxy of GW190425 was not identified, we obtain a SMBH mass-dependent constraint of $r > 7.2 \times (M_\mathrm{BH} / 10^8 M_\odot)^{1/2} \ \mathrm{AU}$ (90\% CL).}
\begin{figure*}[ht!]
	\centering 
	\includegraphics[width=0.9\linewidth]{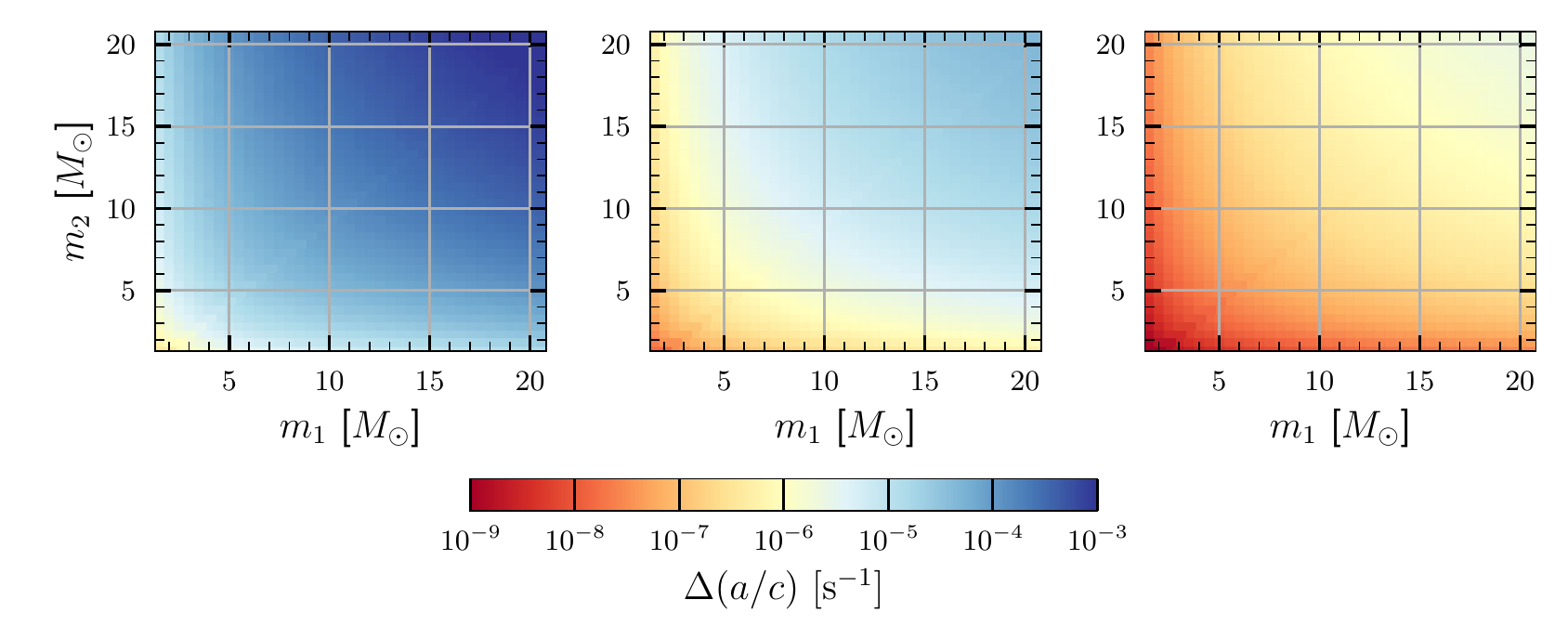}
	\caption{1-$\sigma$  error in the measurement of $ a/c $ for  A+ (left),  CE (middle), ET (right) detector configurations, over a grid of masses and fixed SNR=10.  $ a/c $ is best measured with ET since the low-frequency sensitivity is expected to be better as compared to CE. Not surprisingly, lower masses enable better constraints on $a/c$, because such CBCs spend a longer duration in-band.}
	\label{fig:O53G-constraints}
\end{figure*}
\subsection{Measurement forecasts for future ground-based detector networks}
\label{sec:forecasts-ground-based}
We now outline how the measurement of $ a/c $ will improve with future detectors. We first inject a system into simulated gaussian noise assuming the projected sensitivity of LIGO-Virgo-KAGRA network in the fourth observing run (O4)\footnote{The network contains LIGO detectors at Hanford and Livingston in USA \citep{LIGOScientific:2014pky}, the Virgo detector in Italy \citep{VIRGO:2014yos}, and the KAGRA detector in Japan \citep{KAGRA:2020tym}. T{he corresponding noise power spectral densities are taken from \href{https://dcc.ligo.org/LIGO-T2000012-v1/public}{https://dcc.ligo.org/LIGO-T2000012-v1/public}.}, with component masses similar to GW170817 and three different values of $ a/c = \{0, -10^{-5}, -10^{-3}\} \ \mathrm{s}^{-1}$.   }
The injection template contains the IMRPhenomD~\citep{Khan:2015jqa} waveform approximant with the additional contribution from line-of-sight 
acceleration as in Eq.~\eqref{eq:phase-perturbation}.
We then recover the parameters of the injected signal assuming the full GW phase using Bayesian inference. 
{To probe systematics in recovering the acceleration, we also infer the parameters of the signal with templates containing only the leading-order (i.e. $-4$ PN) term in the acceleration as in Eq. \eqref{eq:phase-perturbation}. The naive expectation would be that neglecting higher-order terms in the phase would incur a higher systematic bias from lower mass systems as compared to high mass systems, just because the phase deviation increases strongly as the mass decreases.}
The recovery $ a/c $ for all GW170817-like injections\footnote{The IMRPhenomD approximant does not include tidal corrections to the phase. Although the non-inclusion of tidal corrections is unphysical for BNS events, we do not expect this choice to affect the posteriors on $a/c$ or the biases due to incomplete terms in the phase.} is shown in the top row of Fig. \ref{fig:O4-constraints}. When using the full phase, all injected values are recovered within the posterior, with the measurement uncertainty being $ \order{10^{-7}} $ $ \mathrm{s}^{-1} $.  {While using the $-4$ PN phase, the injected value is recovered within the posterior for $ a /c = 0, -10^{-5} \ \mathrm{s}^{-1}$, but the recovery is significantly biased when $ a/c = -10^{-3} \ \mathrm{s}^{-1}$}. This illustrates the importance of using an accurate template waveform family while estimating $ a/c $ from BNS events.  We also repeat the same procedure for injection with GW170608-like masses \citep{LIGOScientific:2017vox, LIGOScientific:2018mvr} and plot recoveries in the bottom row of Fig. \ref{fig:O4-constraints}. For all the injected values, recoveries with full phase yield consistent posteriors that include the injected value, with the measurement uncertainty being $ \order{10^{-4}} $ $ \mathrm{s}^{-1} $. The recoveries with $-4$ PN phase are consistent with the full phase recoveries since the higher order correction
has a smaller effect for heavier masses.

In order to forecast constraints for A+ and next-generation (XG) GW detector networks, we resort to a Fisher matrix based approach. Given a frequency-domain GW waveform template $ h(f) $ that depends on a set of parameters $\{\theta_i\}$  in the frequency domain, the elements of the Fisher information matrix $ \Gamma $ can be written as \citep{Cutler:1994ys}:
\begin{equation}\label{key}
	\Gamma_{ij} = \ip{\pdv{h}{\theta_i}}{\pdv{h}{\theta_j}} \qq{,}
\end{equation}
where the inner product $ \ip{a}{b} $ is defined as follows,
\begin{equation}\label{key}
	\ip{a}{b} = 2 \int_{f_\mathrm{min}}^{f_\mathrm{max}} \dd{f} \dfrac{ \qty(a(f) b^*(f) + a^*(f) b(f))}{S_n(f)} \qq{.}
\end{equation}
Here, $ S_n(f) $ stands for the one-sided noise power spectral density (PSD). The covariance matrix $ \Sigma $ of the measurement uncertainties $ \Delta \theta_i $ is the inverse of the Fisher information matrix (i.e. $ \Sigma = \Gamma^{-1} $). The root-mean-square (rms) uncertainty in the measurement of parameter $ \theta_i $ marginalized over all other parameters is given by 	$ \sqrt{\ev{\Delta \theta_i^2}} = \sqrt{\Sigma_{ii}} $.

We use the above prescription to calculate the rms uncertainty in the measurement of $ a/c $. We perform this calculation for three different detector sensitivities:
\begin{enumerate}\label{key}
	 \item LIGO at A+ sensitivity \citep{KAGRA:2013rdx}\footnote{The A+ design PSD was taken from \href{https://dcc.ligo.org/LIGO-T2000012-v1/public}{https://dcc.ligo.org/LIGO-T2000012-v1/public}} with $ f_\mathrm{min} = 15 $ Hz.
	 \item Cosmic Explorer (CE) \citep{CE} at its design 40 km compact-binary optimized sensitivity \citep{2022ApJ...931...22S} with $ f_\mathrm{min} = 5  $ Hz.
	 \item Einstein Telescope (ET) \citep{ET} at its design (ET-D) sensitivity \citep{Hild:2010id} with $ f_\mathrm{min} = 2  $ Hz.
\end{enumerate}
The results for a grid of detector-frame masses at a fixed signal-to-noise ratio (SNR) of $10 $ in each detector are shown in Fig. \ref{fig:O53G-constraints}. For all configurations, we assume that the binary is non-spinning and has no tidal deformability and that the phase without acceleration is modeled by the TaylorF2 approximant including point-particle phase corrections upto 3.5 PN order (see \cite{Buonanno:2009zt} and references therein).  As expected, $ a/c $ is best measured for events with low detector frame mass. It is also evident that the measurement uncertainty is lower for XG detectors as compared to A+. The enhanced low-frequency response of ET results in better constraints as compared to CE. Overall, we find that the best constraint obtained from a 1.4-1.4 $ M_\odot $ binary is $ \order{10^{-9}} \ \qty( \order{10^{-7}} ) \ \mathrm{s}^{-1}$ for XG (A+) detectors, 
while the constraint obtained from a 20-20 $ M_\odot $ binary is $ \order{10^{-5}} \ \qty( \order{10^{-3}} ) \ \mathrm{s}^{-1}$. 

For the constraints above (and the ones that follow in Sec.~\ref{sec:DECIGO}), we do not take into account the effects on the binary waveform produced due to the rotation of the earth. We reiterate that these results are for fixed $\mathrm{SNR}=10$ and that we have verified that the errors scale as $ 1 / \mathrm{SNR} $. As such, we do expect to detect both BNS and BBH events with very high SNRs ($> 100$), especially in XG detectors, and a $ \order{10^{-10}} \ \mathrm{s}^{-1}$ constraint is imminent. 

\begin{figure}[ht!]
	\centering 
	\includegraphics[width=0.9\columnwidth]{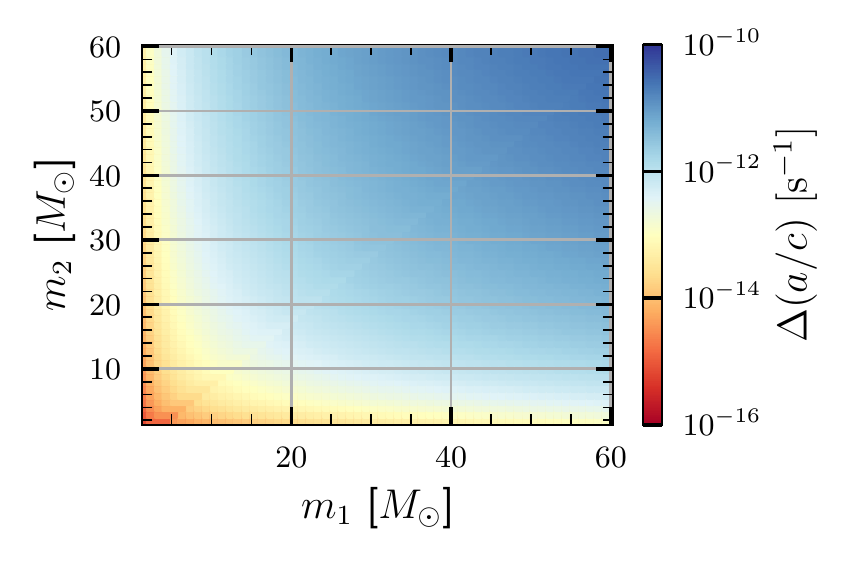}
	\caption{1-$\sigma$ error in the measurement of $ a/c $ in DECIGO over a grid of masses and fixed SNR=10. CBC masses as heavy as $120 M_{\odot}$ still provide constraints of better than $\sim 10^{-11} \ \mathrm{s}^{-1}$. Corresponding precision for BNSs is about $5-6$ orders of magnitude better.}
	\label{fig:DECIGO-constraints}
\end{figure}

\subsection{Measurement forecasts for decihertz detectors}
\label{sec:DECIGO}
Decihertz detectors like DECIGO \citep{Sato:2017dkf} can detect chirping stellar mass binaries in the early stages of inspiral out to very high redshifts. Applying the Fisher matrix formalism,
we calculate the measurement accuracy of $ a/c $ over a grid of masses and fixed $\mathrm{SNR} =10 $, assuming the DECIGO design sensitivity \citep{Yagi:2011wg, PhysRevD.95.109901}  with $ f_\mathrm{min} = 0.1 $ Hz and $ f_\mathrm{max} = 1 $ Hz\footnote{We have assumed that the systems we consider complete a full chirp in the DECIGO band between the assumed minimum and maximum frequency. For the lowest configuration of masses that we consider, the total time in-band is $ \sim 4  $ yrs, comparable to the expected mission duration of a space-based detector.}. The best constraints are $ \order{10^{-16}} $ s$^{-1} $ which are seven orders of magnitude better than the corresponding constraints obtained with audio-band detectors. Even with a 60-60 $M_\odot $ system, the constraints are  $ \order{10^{-10}} $ s$^{-1} $. Again, typical events in DECIGO will have $ \mathrm{SNR} \sim 1000 $, making the best possible constraints with DECIGO  $ \order{10^{-18}} $ s$^{-1}$.

Naively, one would think that the constraints would get better with stellar mass binaries in mHz detectors like LISA~\citep{eLISA} or TianQin~\citep{TianQin:2015yph}. However, most systems in the mass range that we consider would  effectively be monochromatic in such detectors, and their SNRs will also be low. We verified that the constraints here on $ a/c $ are similar to those obtained on $ \dot{G}/G_0 $ in other works \citep{Barbieri:2022zge}.

\section{Summary and Discussion}\label{sec:conclusion}
Recent work \citep{McKernan:2020lgr} has suggested that a significant fraction of LIGO-Virgo's BBHs could have merged in dense stellar environments, including within the disk of AGNs. \cite{Graham:2020gwr} even claim possible evidence of an electromagnetic counterpart to GW190521 produced due to the kick-propelled ejection of this binary BBH merger from an AGN disk. This claim cannot be tested exclusively from the morphology of the observed GW signal due to the relatively large total mass of this BBH and the poor sensitivity of the LVK detectors at low frequencies \footnote{Recent work \citep{Toubiana:2020drf,Sberna:2022qbn} has shown that the AGN provenance of a similar event in LISA could be ascertained from the shape of the waveform.}.

Some works (see, e.g., \cite{Chen:2017xbi}) have even speculated the possibility of mergers in the vicinity of SMBHs. The rate of these mergers is far from constrained, although the current expectation is that more massive CBCs will tend to merge closer to the SMBH than lighter ones, due to mass-segregation. However, other work (see, e.g., \cite{Peng:2021vzr}) have proposed existence of migration traps close to the innermost stable circular orbit of the SMBH which could enable even relatively lighter binaries to merge in the vicinity of the SMBH.

In this work, we study the prospects of constraining line-of-sight acceleration $a/c$ in future observing runs (O4, O5, XG and DECIGO). We then interpret these constraints in the context of mergers around SMBHs, to investigate if constraints on this acceleration could potentially serve as a smoking gun for the provenance of the CBC. We further place the very first GW data-driven constraints on the line-of-sight acceleration for putative BNSs GW170817 and GW190425.

We find that in O4, $a/c$ (in units of $\mathrm{s}^{-1}$) $ = 0, -10^{-5}, -10^{-3}$ can be recovered with a precision of $\sim 1\times 10^{-7}$ at $90\%$ confidence for GW170817-like events. Importantly, however, large accelerations such as $-10^{-3}$ could produce egregious biases in the recovered $a/c$ if only the leading PN order is considered. The precision improves with future observing runs, as would be expected due to increased sensitivity at lower frequencies. BNS-like CBCs will have precisions (assuming $\mathrm{SNR} = 10$) of $\sim  10^{-7} - 10^{-6}$ in O5, $\sim 10^{-8} - 10^{-7}$ in CE (single detector), and $\sim 10^{-9} - 10^{-8}$ in ET (single detector).

\begin{figure*}
	\centering 

 \includegraphics[width=0.8\linewidth]{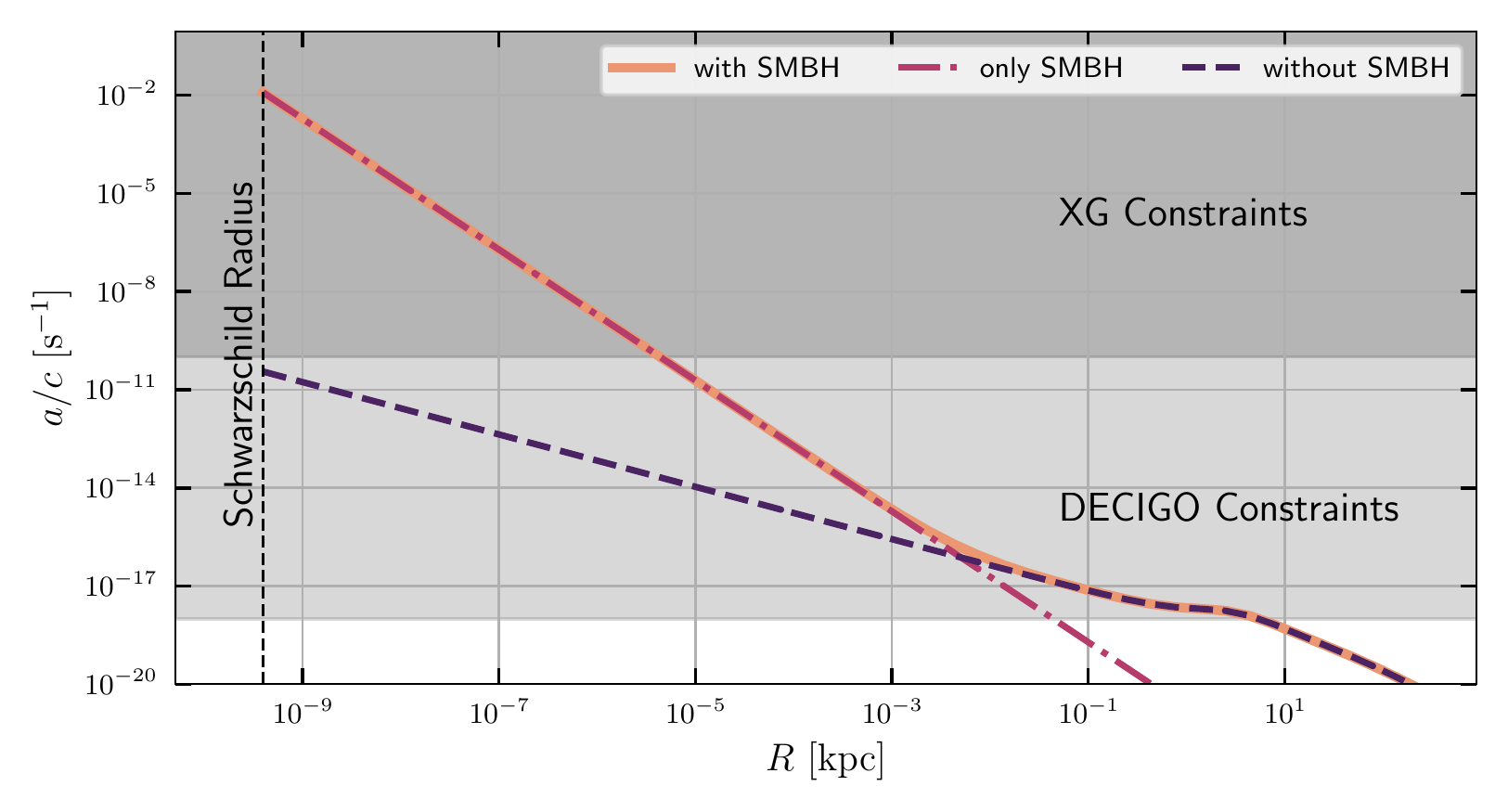}
	\caption{Accelerations in the Milky Way at different distances from its central black hole. The constraints in the figure assume that the acceleration is completely aligned with the line-of-sight, and thus represent a lower limit on the constraints. The accelerations are obtained using the \texttt{MWPotential2014}  potential as defined in \texttt{galpy} \citep{Bovy:2014vfa}, along with a Kepler potential assuming $M_\mathrm{SMBH}=4.154 \times 10^6 M_\odot$ \citep{2019A&A...625L..10G} to account for the central SMBH (orange solid line). The  dash-dotted and dashed lines respectively show accelerations only assuming the Kepler potential and \texttt{MWPotential2014}. {The Schwarzschild radius of the Milky Way SMBH is plotted for reference, along with shaded regions showing constraints obtainable by future detector networks.}}
	\label{fig:milky-way}
\end{figure*}

The low-frequency sensitivity of DECIGO promises spectacular constraints on $a/c$, with precisions that are several orders of magnitude better than XG detectors. Interpreting these constraints in the context of mergers around SMBHs, we find that GW170817-like BNSs could be probed out to distances as large as $R \sim 5 $ kpc from the SMBH in a Milky Way-like galaxy (see Fig.~\ref{fig:milky-way}). In principle, this could enable an investigation of the motion of such BNSs in various parts of the galactic halo outside the SMBH's region of influence\footnote{Note however that derivatives of $a/c$ would also need to be constrained from the GW waveform to infer the position of the binary, and the mass-profile producing the gravitational potential. These can then be used to infer the magnitude of the acceleration of the binary in the potential.}. Even the motion of CBCs with total masses as large as $\mathcal{O}(100 M_{\odot})$ (and $\mathrm{SNR} \sim 1000$) can be probed to a few parsecs from the center of the host galaxy. Such probes would be especially useful to test the claim that a large fraction of BBH mergers reside in AGNs. {The techniques developed in this work can also be extended to probe motion and location of compact binaries in dense stellar environments such as globular clusters or nuclear star clusters} \citep{Inayoshi:2017hgw, Randall:2018lnh, Wong:2019hsq}\footnote{In the context of precision cosmology with DECIGO, measuring the acceleration of the Universe would need to account for the line-of-sight acceleration of the CBCs, both of which appear at the same PN order in the GW phase \citep{Nishizawa:2011eq}.}. 

The non-detection of $a/c$, viz., one that is consistent with $0$ at $90\%$ confidence, can also be used to place constraints on the rate of mergers within a certain distance from the center of host galaxies. This, in turn, could help guide models of mergers in the vicinity of SMBHs, which currently have several uncertainties. We are currently working on sampling of the full GW likelihood (as was done in this work for GW170817 and GW190425) for a host of detected CBCs with total masses less than $20 M_\odot $. We then plan to convert the posteriors on $a/c$ for these events to a posterior on the rate of mergers. We hope to report the results soon.
	
\section*{acknowledgments}
We thank Nathan Johnson-McDaniel and Nicola Tamanini for a careful reading of the draft and constructive comments on the work. Computations were performed on the Alice cluster at ICTS-TIFR and the Sarathi cluster at IUCAA. AV and PA are supported by the Department of Atomic Energy, Government of India, under Project No. RTI4001. AV is also supported by a Fulbright Program grant under the Fulbright-Nehru Doctoral Research Fellowship, sponsored by the Bureau of Educational and Cultural Affairs of the United States Department of State and administered by the Institute of International Education and the United States-India Educational Foundation. K.G.A. acknowledges the Swarnajayanti grant DST/SJF/PSA-01/2017-18 of the Department of Science and Technology, India and support from Infosys Foundation.

This research has made use of data or software obtained from the Gravitational Wave Open Science Center (gwosc.org), a service of LIGO Laboratory, the LIGO Scientific Collaboration, the Virgo Collaboration, and KAGRA \citep{LIGOScientific:2019lzm}. LIGO Laboratory and Advanced LIGO are funded by the United States National Science Foundation (NSF) as well as the Science and Technology Facilities Council (STFC) of the United Kingdom, the Max-Planck-Society (MPS), and the State of Niedersachsen/Germany for support of the construction of Advanced LIGO and construction and operation of the GEO600 detector. Additional support for Advanced LIGO was provided by the Australian Research Council. Virgo is funded, through the European Gravitational Observatory (EGO), by the French Centre National de Recherche Scientifique (CNRS), the Italian Istituto Nazionale di Fisica Nucleare (INFN), and the Dutch Nikhef, with contributions by institutions from Belgium, Germany, Greece, Hungary, Ireland, Japan, Monaco, Poland, Portugal, Spain. KAGRA is supported by the Ministry of Education, Culture, Sports, Science and Technology (MEXT), Japan Society for the Promotion of Science (JSPS) in Japan; National Research Foundation (NRF) and the Ministry of Science and ICT (MSIT) in Korea; Academia Sinica (AS) and National Science and Technology Council (NSTC) in Taiwan.

\textit{Software}: \texttt{NumPy} \citep{vanderWalt:2011bqk}, \texttt{SciPy} \citep{Virtanen:2019joe}, \texttt{astropy} \citep{2013A&A...558A..33A, 2018AJ....156..123A}, \texttt{Matplotlib} \citep{Hunter:2007}, \texttt{jupyter} \citep{jupyter}, \texttt{pandas} \citep{mckinney-proc-scipy-2010}, \texttt{galpy} \citep{Bovy:2014vfa}, \texttt{dynesty} \citep{Speagle:2019ivv}, \texttt{bilby} \citep{Ashton:2018jfp} and \texttt{PESummary} \citep{Hoy:2020vys}.

\bibliographystyle{aasjournal}
\bibliography{references}

\appendix
\section{Derivation of the GW phase and amplitude correction}
\label{sec:appendix}
Here we describe the steps required to compute the corrections to the phase and amplitude of the Fourier domain waveform due to the time-dependent Doppler shift caused by the line-of-sight acceleration of the binary. While the original waveforms (without the Doppler shift) are generated using either the {IMRPhenomD} or {IMRPhenomPv2\_NRTidal} approximant, we use the non-spinning 3.5PN waveforms to compute the Doppler shift, as we expect their accuracy to be sufficient for our purpose. 

Let $z_l = \Gamma t_o $ be the (time-dependent) redshift due to the line-of-sight acceleration of the binary of total mass $M$, where $\Gamma = a / c$, $a$ is the line-of-sight acceleration, and $t_o$ is the observation time. Hereon, we will be working in $G = c = 1$ units. Let us further define $v_u = (\pi M f_u)^{1/3}$, where $f_u$ is the unperturbed Fourier frequency, and let $f_o$ be the Doppler shifted Fourier frequency. 

Then, for $\Gamma t_o << 1$,
\begin{align}
    \label{A1}
        f_u &= f_o (1 + z_l) \qq{:} \mathrm{redshift},\\
        dt_u &= dt_o/(1+z_l) \qq{:} \mathrm{time\,dilation}
        \\
        v_u &= v_o (1 + z_l)^{1/3}
\end{align}
where $v_o = (\pi M f_o)^{1/3}$. We can further write:
\begin{equation}
    \label{A2}
    \frac{dv_u}{dt_u} = (1 + \Gamma t_o)^{4/3}\left( \frac{dv_o}{dt_o} + \frac{\Gamma v_o}{3} \right) \quad \Rightarrow \quad  \frac{dv_o}{dt_o} = - \frac{\Gamma v_o}{3} + (1 + \Gamma t_o)^{-4/3}\frac{dv_u}{dt_u}
\end{equation}
\par Under the stationary phase approximation (SPA), $\frac{dv_u}{dt_u}$ is given by Eq. (3.6) of \citep{Buonanno:2009zt} with $v$ replaced by $v_u$ and $\nu$ by the symmetric mass ratio $\eta$. Substituting this relation in Eq. (\ref{A2}) together with $v_u$ from Eq. (\ref{A1}), we get $\frac{dv_o}{dt_o}$ in terms of $v_o$ and $z_l$. Note that $z_l$ still has a factor of $t_o$. Hence as a next step, we use the relationship between $t_o$ and $v_o$, which is given by\footnote{The reason behind using $v_o$ instead of $v_u$ is the presence of $\Gamma$ with $t_o$ in $z_l$ and using $v_u$ would have made it second order in $\Gamma$.} Eq. (3.8b) of \citep{Buonanno:2009zt}, to get $\frac{dv_o}{dt_o}$ in terms of $v_o$ and $\Gamma$. In the final step, we invert this equation to get $\frac{dt_o}{dv_o}$. 

Integrating $\frac{dt_o}{dv_o}$ in the limits $v_o \to v_{lso} \equiv (\pi M f_{lso})^{1/3}$ and $v_o \to v \equiv (\pi M f)^{1/3}$, where $f_{lso} = \frac{1}{6^{3/2} \pi M}$ and $f$ is the observed frequency of the GW, we get (in geometrized units): 
\begin{multline}
\label{A3}
t - t_c = - \frac{5M}{256 \eta} v^{-8} \Biggl[1 + \left(\frac{743}{252} + \frac{11}{3}\eta  \right)v^2 - \frac{32\pi}{5}v^3 + 2 \left( \frac{3058673}{1016064} + \frac{5429}{1008}\eta + \frac{617}{144}\eta^2 \right) v^4 - \left( \frac{7729}{252} - \frac{13}{3}\eta \right) \pi v^5 \\ + \Biggl\{ - \frac{10052469856691}{23471078400} + \frac{128}{3}\pi^2 + \frac{6848}{105}\gamma  + \left(\frac{3147553127}{3048192} - \frac{451}{12}\pi^2 \right)\eta  - \frac{15211}{1728}\eta^2 + \frac{25565}{1296}\eta^3 + \frac{6848}{105}\ln (4v) \Biggr\}v^6 \\ - \left(\frac{15419335}{127008} + \frac{75703}{756}\eta - \frac{14809}{378}\eta^2 \right) \pi v^7 + \frac{65\Gamma M}{1536 \eta}  v^{-8} \Biggl\{1 + \frac{22}{13} \left(\frac{743}{252} + \frac{11}{3}\eta  \right) v^2 -  \frac{128}{13} \pi v^3 + \Biggl( \frac{1755623}{122304} + \frac{32633}{1092}\eta \\ + \frac{1101}{52}\eta^2 \Biggr) v^4 - \left( \frac{83228}{1365} + \frac{4592}{195}\eta \right) \pi v^5 + \Biggl\{ - \frac{2274117187691}{5029516800} + \frac{66304}{975}\pi^2 + \frac{13696}{195}\gamma + \left( \frac{3311653861}{2830464} - \frac{3157}{78}\pi^2 \right)\eta \\ + \frac{2030687}{33696}\eta^2 + \frac{35693}{648}\eta^3 + \frac{13696}{195}\ln (4v) \Biggr\} v^6  - \left(\frac{158992529}{687960} + \frac{1015907}{4095}\eta - \frac{838}{4095}\eta^2 \right) \pi v^7 \Biggr\} \Biggr] 
\end{multline}
where all of the terms containing $v_{lso}$ are just constants and have been absorbed in $t_c$ (time at the coalescence).
\par Observing that the infinitesimal orbital phase $d\phi$ remains invariant, we can rewrite Eq. (3.3a) of \citep{Buonanno:2009zt} as $d\phi = \frac{v_o^3}{M} \frac{dt_o}{dv_o}dv_o$. Integrating this in the limits same as before, we get:
\begin{multline}
\label{A4}
\phi(f) = \frac{\phi_c}{2} - \frac{v^{-5}}{32\eta} \Biggl[1 + \left(\frac{3715}{1008} + \frac{55}{12}\eta  \right)v^2 - 10 \pi v^3 + 5 \left( \frac{3058673}{1016064} + \frac{5429}{1008}\eta + \frac{617}{144}\eta^2 \right) v^4 + \left( \frac{38645}{672} - \frac{65}{8}\eta \right) \pi v^5 \ln \left( \frac{v}{v_{lso}} \right) \\ - \Biggl\{ - \frac{12348611926451}{18776862720} + \frac{160}{3}\pi^2 + \frac{1712}{21}\gamma  + \left(\frac{15737765635}{12192786} - \frac{2255}{48}\pi^2 \right)\eta - \frac{76055}{6912}\eta^2 + \frac{127825}{5184}\eta^3 + \frac{1712}{21}\ln (4v) \Biggr\}v^6 \\ + \left(\frac{77096675}{2032128} + \frac{378515}{12096}\eta - \frac{74045}{6048}\eta^2 \right) \pi v^7 + \frac{25\Gamma M}{768\eta} v^{-8} \Biggl\{1 + \frac{7}{3} \left(\frac{743}{336} + \frac{11}{4}\eta  \right) v^2 -  \frac{52}{5} \pi v^3 + \Biggl( \frac{1755623}{112896} + \frac{32633}{1008}\eta \\ + \frac{367}{16}\eta^2 \Biggr) v^4 - \left( \frac{228877}{3360} + \frac{3157}{120}\eta \right) \pi v^5 + \Biggl\{ - \frac{5873342252515}{11266117632} + \frac{1184}{15}\pi^2 + \frac{1712}{21}\gamma + \left( \frac{16558269305}{12192768} - \frac{2255}{48}\pi^2 \right)\eta \\ + \frac{10153435}{145152}\eta^2 + \frac{331435}{5184}\eta^3 + \frac{1712}{21}\ln (4v) \Biggr\} v^6 - \left(\frac{158992529}{564480} + \frac{1015907}{3360}\eta - \frac{419}{1680}\eta^2 \right) \pi v^7 \Biggr\} \Biggr]
\end{multline}
where $\phi_c$ is the phase at the coalescence. Here all of the terms containing $v_{lso}$ have been absorbed in $\phi_c$ except the $\log$ term. 
\par Substituting Eq. (\ref{A3}) and Eq. (\ref{A4}) in $\Psi(f) = -\pi/4 + 2 \pi f t(f) - \phi(f)$, we get the total phase.
The correction in the phase is then simply $(\Psi_{acc}(f) - \Psi_{no\,acc}(f) \equiv \Psi(f)-\Psi_{3.5}(f))$, which is given by:
\begin{multline}
\label{A6}
\Delta \Psi(f) = \frac{25\Gamma M}{65536\eta^2} v^{-13} \Biggl[ 1 + \left(\frac{743}{126} + \frac{22}{3}\eta  \right) v^2  - \frac{64 \pi}{5} v^3 + \left( \frac{1755623}{84672} + \frac{32633}{756}\eta + \frac{367}{12}\eta^2 \right) v^4 \\ - \left( \frac{20807}{210} + \frac{574}{15}\eta \right) \pi v^5 +  \Biggl\{ - \frac{28907482848623}{35206617600} + \frac{9472}{75}\pi^2 + \frac{13696}{105}\gamma + \frac{13696}{105}\ln (4v) + \Biggl( \frac{3311653861}{1524096} \\ - \frac{451}{6}\pi^2 \Biggr)\eta  + \frac{2030687}{18144}\eta^2 + \frac{66287}{648}\eta^3 \Biggr\}v^6 - \Biggl( \frac{158992529}{317520} + \frac{1015907}{1890}\eta - \frac{419}{945}\eta^2 \Biggr)\pi v^7 \Biggr]
\end{multline}

\par Assuming $\iota = 0$ in Eq. (4.361) of \citep{maggiore2007gravitational} and observing that $\frac{d \Phi}{d t_o} = \omega_{GW} = 2 \pi f_o = \frac{2v_o^3}{M} > 0$ i.e. $\frac{d^2 \Phi}{d t_o^2} =  \frac{6 v_o^2}{M}\frac{d v_o}{d t_o} > 0$, we find, using the Eq. (4.366) of the same, the amplitude correction \footnote{We make the substitution $v_o \equiv v$ since we have already assumed the SPA.} (considering the corrections only to the Newtonian order) to be given by:

\begin{equation}
    \label{A7}
    \frac{\mathcal{A}}{\mathcal{A}_{lead}} = 1 + \frac{65 \Gamma  M}{2048 \eta  v^8} [ 1 + \mathcal{O}(v^2)]
\end{equation}

where $\mathcal{A}_{lead}$ is the leading order amplitude and is given by the Eq. (4.369) of the same reference with $\iota = 0$. We have checked that the amplitude correction does not affect our constraints on $\Gamma$ significantly enough to merit inclusion.

\end{document}